\documentclass[11pt,a4paper,twocolumn]{article}

\usepackage[utf8]{inputenc}
\usepackage[english]{babel}
\usepackage{caption}
\usepackage{amsmath}
\usepackage{amsfonts}
\usepackage{amssymb}
\usepackage{amsthm}
\usepackage{graphicx}
\usepackage{verbatim}
\usepackage{psfrag}
\usepackage{geometry}
\usepackage{enumerate}
\usepackage{fancyhdr}
\usepackage[auth-lg]{authblk}
\usepackage[semicolon,comma]{natbib}

\title{\Large \bf Representation of $\mathfrak{su}(8)$ in Pauli basis}
\author[1*]{K. Y. Chew}
\author[1,2*]{Nurisya M. Shah}
\author[1,2]{K. T. Chan}
\affil[1]{Laboratory of Computational Sciences and Mathematical Physics, Institute for Mathematical Research (INSPEM), Universiti Putra Malaysia, Malaysia}
\affil[2]{Department of Physics, Faculty of Science, Universiti Putra Malaysia, 43400 UPM Serdang, Selangor, Malaysia}
\date{\vspace{-1\baselineskip}{\textit{$^*$Corresponding author: gs52076@student.upm.edu.my, risya@upm.edu.my}}}

\fancypagestyle{plain}{%
\fancyhf{} 
\fancyhead[C]{\small{}}
\fancyfoot[C]{\small{\thepage}}

}

\fancypagestyle{fancy2}{
\fancyhf{} 
\lhead{\small}

}

\begin{document}
\renewcommand{\abstractname}{\vspace{-3\baselineskip}}
\pagestyle{plain}
\twocolumn[
    \begin{@twocolumnfalse}
    \maketitle
	\begin{abstract} \noindent
		Quantum computation started to become significant field of studies as it hold great promising towards the upgrade of our current computational power. Studying the evolution of quantum states serves as a good fundamental in understanding quantum information which lead to quantum computation. This was assisted with the respective mathematical tools such as Lie group and Lie algebra. In this study, the Lie algebra of $\mathfrak{su}(8)$ is represented in tensor product between three Pauli matrices. This is done by constructing the generalized Gell-Mann matrices and compared to the Pauli basis. This study will explicitly shows the one-to-one correlation of Gell-Mann matrices with the Pauli basis resembled change of coordinates. This is particularly useful when dealing with quantum circuit problems.
	\end{abstract}
	\end{@twocolumnfalse}
	]

\thispagestyle{fancy2}

\section{Introduction}\label{introduction}
Studies of quantum mechanics have became a staple foundation to have a good grasp on the fundamentals of nature in order to harness those power for future technologies. Some of the promising applications are implementation of Shor's algorithm for prime factorization problem \citep{shor_1994}, and implementation of Grover's search algorithm together with genetic algorithm for determining graph's planarity \citep{grover_1996}, \citep{kureychik_2019}. One of the important studies to apply quantum mechanics into physically realizable system is by using group theory. As an example, observables such as energy, $E$ can only be obtained from the application of operators in $SO(n)$ group to the quantum states as it ensure the eigenvalues to be real, while evolution of quantum systems needed to be governed by operators in $SU(n)$ group to preserved the unit probability of the systems. Group theory also have wide application in physics, chemistry, and material sciences due to groups preserving the symmetry properties of the system which leads to invariant property. Some of the simple objects such as a circle arise naturally from group theory in particularly $e^{i\theta}\in U(1)$ where $\theta$ traced how the path of the circles. This concept was eventually extended to Lie group by Sophus Lie in 1884 to encompass ideas of the application of group theory to geometrical objects. This enable Lie group to have both group and geometry properties, which is particularly useful as certain mathematics problems in group theory can be recast into geometry or vice versa for different insights \citep{Brian2015}. 

Alongside with Lie group, Lie algebra is the derivative of Lie group, thus linearizing the problems of study interest. Lie algebra is one of the representation of its Lie group, this simplify the calculation and computation of the problems. Since Lie algebra is a linear object, it could be constructed by linearly combining its basis. In particular $\mathfrak{su}(n)$ Lie algebra is constructed by the corresponding generalized Gell-Mann matrices. However, the choice of any basis is just the preferences of the user. Thus it would be possible to construct (or represent) $\mathfrak{su}(n)$ in other basis, in our case in terms of tensor product of Pauli matrices (or Pauli coordinates) for $\mathfrak{su}(2^n)$. The motivation behind this representation is due to (1) it is much easier to compute tensor product of Pauli matrices as compared to generalized Gell-Mann matrices, (2) for certain problems (quantum circuit problems), having a Pauli coordinates give better insights of the evolution of states contributed by quantum gates. This is particularly useful in for representing the evolution of the quantum states using geometrical tools, by representing the Hamiltonian of the evolution of the quantum states in terms of Pauli coordinates \citep{Nielsen_2006}, \citep{Dowling_2008}, \citep{Brandt_2009}.

To my knowledge, Lie algebra of $d$-dimensional qudits system ($\mathfrak{su}(d)$) can be decomposed by either three different matrix bases in particular the generalized Gell-Mann matrix basis, the polarization operator basis and the Weyl operator basis  \citep{Bertlmann_2008}. On the same page, an $n$-qubit system, resembled a $2^n$-dimensional qudits system have its Lie algebra $\mathfrak{su}(2^n)$, thus could be decomposed too by the above matrix bases. In particular for an $n$-qubit system it could also be decomposed by tensor product of Pauli matrices, which is explicitly shown in this work despite it is a well known ideas. And it is shown to be one-to-one corresponded or just a change of basis to the generalized Gell-Mann matrices. In line with motivation (2), we therefore prepare this work such that, proper evolution of quantum states in terms of Pauli coordinates will be properly understood. \\

This paper is organized as follows. Section 2 describe the general form for $SU(n)$ Lie group namely $SU(8)$ and provide the counting method for the number of free parameters for $SU(8)$. In section 3, the generator for SU(8) is constructed, denoted by $\mathfrak{su}(8)$ Lie algebra having a well-known form to be the Gell-Mann matrices. In section 4, the tensor product of Pauli matrices is computed and they are categorize according to their different forms. Relation between generators of $SU(8)$ with tensor product of Pauli matrices associated to their forms will be given in section 5. In section 6, results from section 5 is shown explicitly highlighting the concept of change of basis. The final section shall offers some discussion and conclusions.
\section{General form for SU(n)}
Let us start by constructing a general form for $SU(8)$ Lie group. For any element $SU(8)$, it should satisfies: 
\begin{enumerate}[(i)]
	\itemsep-0.3em
	\item The diagonals must be real.
	\item The diagonals must be traceless. 
	\item The upper triangle of the matrix must be the complex conjugate of the lower triangle of the matrix. 
\end{enumerate}
These condition is imposed by the Hermiticity of $SU(n)$.  \\
With that condition, the number of free parameter can be calculated:
\begin{enumerate}[(i)]
	\itemsep-0.3em
	\item There is $2n^2$ free parameter due to $n\times n$ complex entries of $SU(n)$ matrix,
	\item The $n$ diagonals must be real thus removing $n$ free parameter,
	\item The upper triangle of the matrix must be the complex conjugate of the lower triangle of the matrix, the upper triangle having $2(n^2-n)/2$ components, where $n^2-n$ is obtained by removing the $n$ diagonals from $n^2$ components of matrix, multiplied by 2 due to complex component and divided by 2 to remove the lower triangle component. With this condition we removed $2(n^2-n)/2$ components. 
	\item By applying step (i), (ii) and (iii) we obtained
	\begin{equation*}
	2n^2-n-\frac{2(n^2-n)}{2}=n^2.
	\end{equation*}
	\item Finally due to ``special" properties of SU(8) there is an additional restrains which is $\det(SU(n))=1$ thus removing 1 more free parameter yielding $n^2-1$ free parameter. 
\end{enumerate}
Given our case are $SU(8)$ we have $8^2-1=63$ free parameter.\\
\\
Thus give rise of the below form. \\
$G=\\
{\small\begin{pmatrix}
\psi_1 & a_1-ib_1 &  \dots & a_6-ib_6 & a_7-ib_7 \\
a_1+ib_1 & \psi_2 & \dots & a_{12}-ib_{12} & a_{13}-ib_{13} \\
\vdots & \vdots & \ddots & \vdots &\vdots \\
a_6+ib_6 & a_{12}+ib_{12} & \dots & \psi_7 & a_{28}-ib_{28} \\
a_7+ib_7 & a_{13}+ib_{13} & \dots & a_{28}+ib_{28} & \psi_8
\end{pmatrix}}$\\
\\
where $G\in SU(8)$.\\
Due to the ``special" condition of $SU(8)$ which lead to traceless $G$, $\psi_8=-\psi_1-\psi_2-\dots-\psi_7$ is required.\\
With that, we have $\psi_1$ to $\psi_7$, 7 parameter, $a_1$ to $a_{28}$, 28 parameter, $b_1$ to $b_{28}$, 28 parameter, lead to a total of 7+28+28=63 free parameters.

\section{Generator for $SU(8)$ Lie Group}
Let $G$ be a matrix Lie group.  Then the \textbf{Lie algebra} of $G$ denoted $\mathfrak{g}$, is the set of all matrices $X$ such that $e^{tX}$ is in $G$ for all real numbers $t$ \citep{Brian2015}. \\
Given:
\begin{equation*}
\frac{d}{dt}e^{tX}=Xe^{tX}=e^{tX}X, \quad \frac{d}{dt}\bigg|_{t=0}e^{tX}=X.
\end{equation*}
We can obtained the generator, $X$. \\
For the case $SU(8)$, its generators is the element of $\mathfrak{su}(8)$ where $G\in SU(8)$ is generated by linear combination of above 63 free parameters thus having 63 generators, labeled as $X_i\in\mathfrak{su}(8)$, where $i=1,2,\dots,63$. Thus $G$ can be written as:
\begin{align*}
	G=\exp&(\psi_1X_1+\dots+\psi_7X_7\\
	&+a_1X_8+\dots+a_{28}X_{35}\\
	&+b_1X_{36}+\dots+b_{28}X_{63}).
\end{align*}
Following up that, the 63 generators can be found,\\
for the diagonal part:
\begin{gather*}
\frac{dG}{d\psi_1}\bigg|_{\psi_1=\dots=\psi_7=a_1=\dots=a_{28}=b_1=\dots=b_{28}=0}=X_1,\\
\vdots\\
\frac{dG}{d\psi_7}\bigg|_{\psi_1=\dots=\psi_7=a_1=\dots=a_{28}=b_1=\dots=b_{28}=0}=X_7,\\
\end{gather*}
for the off-diagonal real parts:
\begin{gather*}
	\frac{dG}{da_1}\bigg|_{\psi_1=\dots=\psi_7=a_1=\dots=a_{28}=b_1=\dots=b_{28}=0}=X_8,\\
	{\centering\vdots}\\
	\frac{dG}{da_{28}}\bigg|_{\psi_1=\dots=\psi_7=a_1=\dots=a_{28}=b_1=\dots=b_{28}=0}=X_{35},\\
\end{gather*}
and for the off-diagonal imaginary parts:
\begin{gather*}
\frac{dG}{db_1}\bigg|_{\psi_1=\dots=\psi_7=a_1=\dots=a_{28}=b_1=\dots=b_{28}=0}=X_{36,}\\
\vdots\\
\frac{dG}{db_{28}}\bigg|_{\psi_1=\dots=\psi_7=a_1=\dots=a_{28}=b_1=\dots=b_{28}=0}=X_{63},\\
\end{gather*}
where $X_i$, $i=8,9,\dots ,63$ is the generalized Gell-Mann matrices for SU(8), while $X_i$, $i=1,2,\dots 7$ can be linearly combined to obtain the diagonal part for the generalized Gell-Mann matrices for SU(8).

\section{Tensor Product of Pauli Matrices}
Pauli Matrices can be categorized in 2 groups which are the diagonals (labeled as $D$):
\begin{equation*}
I=
\begin{pmatrix}
1 & 0\\
0 & 1
\end{pmatrix},\quad
\sigma_z=
\begin{pmatrix}
1 & 0\\
0 & -1
\end{pmatrix},
\end{equation*}
and the off-diagonals (labeled as $OD$):
\begin{equation*}
\sigma_x=
\begin{pmatrix}
0 & 1\\
1 & 0
\end{pmatrix},\quad
\sigma_y=
\begin{pmatrix}
0 & -i\\
i & 0
\end{pmatrix},
\end{equation*}
Thus the tensor product of Pauli matrices can come in 8 forms by permutation as follows:
\begin{align*}
	D\otimes D\otimes D, &\quad  D\otimes D\otimes OD,\\
	D\otimes OD\otimes D, &\quad  D\otimes OD\otimes OD,\\
	OD\otimes D\otimes D, &\quad  OD\otimes D\otimes OD,\\
	OD\otimes OD\otimes D, &\quad  OD\otimes OD\otimes OD.\\
\end{align*}
For example, $ D\otimes D\otimes OD$:
\begin{equation*}
	\begin{pmatrix}
	0 & x & 0 & 0 & 0 & 0 & 0 & 0 \\
	x & 0 & 0 & 0 & 0 & 0 & 0 & 0 \\
	0 & 0 & 0 & x & 0 & 0 & 0 & 0 \\
	0 & 0 & x & 0 & 0 & 0 & 0 & 0 \\
	0 & 0 & 0 & 0 & 0 & x & 0 & 0 \\
	0 & 0 & 0 & 0 & x & 0 & 0 & 0 \\
	0 & 0 & 0 & 0 & 0 & 0 & 0 & x \\
	0 & 0 & 0 & 0 & 0 & 0 & x & 0 \\
	\end{pmatrix},
\end{equation*}
where $x\in\{1,i,-1,-i\}$ being the quaternion.\\
And one of the element from $ D\otimes D\otimes OD$ form is as followed:
\begin{align*}
	\sigma_z\otimes I\otimes \sigma_y 
	= \\
	\begin{pmatrix}
	0 & -i & 0 & 0 & 0 & 0 & 0 & 0 \\
	i & 0 & 0 & 0 & 0 & 0 & 0 & 0 \\
	0 & 0 & 0 & -i & 0 & 0 & 0 & 0 \\
	0 & 0 & i & 0 & 0 & 0 & 0 & 0 \\
	0 & 0 & 0 & 0 & 0 & i & 0 & 0 \\
	0 & 0 & 0 & 0 & -i & 0 & 0 & 0 \\
	0 & 0 & 0 & 0 & 0 & 0 & 0 & i \\
	0 & 0 & 0 & 0 & 0 & 0 & -i & 0 \\
	\end{pmatrix}.&
\end{align*}

\section{Representing Generator of SU(8) in Pauli Matrices}
The generator can be represented in the categories of form in the previous section. The results is as follows:$\vspace{0.2cm}$\\
$D\otimes D\otimes D$ form generator:\\
 $X_1,X_2,X_3,X_4,X_5,X_6,X_7$ $\vspace{0.2cm}$\\
$D\otimes D\otimes OD$ form generator:\\
Real part : $X_8,X_{21}, X_{30}, X_{35}$\\
Imaginary part : $X_{36},X_{49}, X_{58}, X_{63}$ $\vspace{0.2cm}$\\
$D\otimes OD\otimes D$ form generator:\\
Real part : $X_9,X_{16}, X_{31}, X_{34}$\\
Imaginary part : $X_{37},X_{44}, X_{59}, X_{62}$ $\vspace{0.2cm}$\\
$D\otimes OD\otimes OD$ form generator:\\
Real part : $X_{10},X_{15}, X_{32}, X_{33}$\\
Imaginary part : $X_{38},X_{43}, X_{60}, X_{61}$ $\vspace{0.2cm}$\\
$OD\otimes D\otimes D$ form generator:\\
Real part : $X_{11},X_{18}, X_{24}, X_{29}$\\
Imaginary part : $X_{39},X_{46}, X_{52}, X_{57}$ $\vspace{0.2cm}$\\
$OD\otimes D\otimes OD$ form generator:\\
Real part : $X_{12},X_{17}, X_{25}, X_{28}$\\
Imaginary part : $X_{40},X_{45}, X_{53}, X_{56}$ $\vspace{0.2cm}$\\
$OD\otimes OD\otimes D$ form generator:\\
Real part : $X_{13},X_{20}, X_{22}, X_{27}$\\
Imaginary part : $X_{41},X_{48}, X_{50}, X_{55}$ $\vspace{0.2cm}$\\
$OD\otimes OD\otimes OD$ form generator:\\
Real part : $X_{14},X_{19}, X_{23}, X_{26}$\\
Imaginary part : $X_{42},X_{47}, X_{51}, X_{54}$ $\vspace{0.2cm}$\\
As an example:
\begin{align*}
	X_1=\frac{1}{4}&\big[(I\otimes I \otimes \sigma_z)+(\sigma_z\otimes I \otimes I)\\
	&+(I\otimes \sigma_z \otimes I)+ (\sigma_z\otimes \sigma_z \otimes\sigma_z)\big].
\end{align*}

\section{Change of Basis from Generator Basis to Pauli Basis}
The result above shown to have one-to-one correspondent between generator of $SU(8)$ Lie group and tensor product of Pauli matrices. According to change of basis equation:
\begin{equation*}
	y^\mu=g^\mu_\nu x^\nu,
\end{equation*}
where $y^\mu, x^\nu, \mu,\nu=1,2,3,4$ are different Lie algebra basis, $g_{\mu\nu}$ are the transformation matrix,
Given our both $H$ and $J$ bases transform orthogonally, one can get away without worrying about the distinction between covariant and contravariant part of the tensor, and also it satisfied the properties of $(g^\mu_\nu)^{-1}=(g^\mu_\nu)^T=g^\nu_\mu$. Let us look at this example:\\ 
$D\otimes D\otimes OD$ form generator:\\
Real part :
\begin{align*}
I \otimes I \otimes \sigma_x &= X_8 + X_{21} + X_{30} + X_{35},\\
I \otimes \sigma_z \otimes \sigma_x &= X_8 - X_{21} + X_{30} - X_{35},\\
\sigma_z \otimes I \otimes \sigma_x &= X_8 + X_{21} - X_{30} - X_{35},\\
\sigma_z \otimes \sigma_z \otimes \sigma_x &= X_8 - X_{21} - X_{30} + X_{35}.   
\end{align*}
Let:
\begin{align*}
	y^1=I \otimes I \otimes \sigma_x, \hspace{0.72cm} & x^1= X_8,\\
	y^2=I \otimes \sigma_z \otimes \sigma_x, \hspace{0.50cm} & x^2=X_{21},\\
	y^3=\sigma_z \otimes I \otimes \sigma_x, \hspace{0.50cm} & x^3=X_{30},\\
	y^4=\sigma_z \otimes \sigma_z \otimes \sigma_x, \hspace{0.3cm} & x_4=X_{35},   
\end{align*}
one can rewrite this using the equation above:
\begin{align*}
	y^\mu&=g^\mu_\nu x^\nu,\\
	\begin{pmatrix}
	y^1\\
	y^2\\
	y^3\\
	y^4
	\end{pmatrix}
	&=
	\begin{pmatrix}
	1 & 1 & 1 & 1\\
	1 & -1 & 1 & -1\\
	1 & 1 & -1 & -1\\
	1 & -1 & -1 & 1
	\end{pmatrix}
	\begin{pmatrix}
	x^1\\
	x^2\\
	x^3\\
	x^4
	\end{pmatrix}.
\end{align*}
By finding the inverse of $g^{\mu}_{\nu}$ one can change the basis of $x^\nu$ which is the generator of $SU(8)$ back to tensor product of Pauli matrices. 
\begin{align*}
(g^\mu_\nu)^{-1} y^\mu&=(g^\mu_\nu)^{-1}g^\mu_\nu x^\nu,\\
x^\nu&=(g^\mu_\nu)^{-1} y^\mu,\\
x^\nu&=g^\nu_\mu y^\mu,\\
\begin{pmatrix}
x^1\\
x^2\\
x^3\\
x^4
\end{pmatrix}
&=\frac{1}{4}
\begin{pmatrix}
1 & 1 & 1 & 1\\
1 & -1 & 1 & -1\\
1 & 1 & -1 & -1\\
1 & -1 & -1 & 1
\end{pmatrix}
\begin{pmatrix}
y^1\\
y^2\\
y^3\\
y^4
\end{pmatrix},
\end{align*}
which is as followed
\begin{align*}
	X_{8}= \frac{1}{4}\Big[&(I\otimes I\otimes \sigma_x)+(I\otimes \sigma_z \otimes \sigma_x)\\&+(\sigma_z \otimes I\otimes \sigma_x)+(\sigma_z\otimes \sigma_z \otimes \sigma_x)\Big],\\
	X_{21}= \frac{1}{4}\Big[&(I\otimes I\otimes \sigma_x)-(I\otimes \sigma_z \otimes \sigma_x)\\&+(\sigma_z \otimes I\otimes \sigma_x)- (\sigma_z\otimes \sigma_z \otimes \sigma_x)\Big],\\
	X_{30}= \frac{1}{4}\Big[&(I\otimes I\otimes \sigma_x)+(I\otimes \sigma_z \otimes \sigma_x)\\&-(\sigma_z \otimes I\otimes \sigma_x)- (\sigma_z\otimes \sigma_z \otimes \sigma_x)\Big],\\
	X_{35}= \frac{1}{4}\Big[&(I\otimes I\otimes \sigma_x)-(I\otimes \sigma_z \otimes \sigma_x)\\&-(\sigma_z \otimes I\otimes \sigma_x)+ (\sigma_z\otimes \sigma_z \otimes \sigma_x)\Big].\\
\end{align*}

Note that $(g^\mu_\nu)^{-1}=g^\nu_\mu$, when:
\begin{equation*}
g^\mu_\nu = \frac{1}{2}
	\begin{pmatrix}
	1 & 1 & 1 & 1\\
	1 & -1 & 1 & -1\\
	1 & 1 & -1 & -1\\
	1 & -1 & -1 & 1
	\end{pmatrix}= g^\nu_\mu,
	\end{equation*} 
which is different than the one before by a factor of $\frac{1}{2}$. Note that, this is still acceptable as the basis are normalized unit, thus changing the magnitude would not affects the properties of a basis. 

\section{Discussion and Conclusion}
From the result, it is shown that all 63 generators of $SU(8)$ Lie group can be represented in linear combination of the tensor product of Pauli matrices. All 63 generators have unique linear combination which is one-to-one correspondent similar to change of basis, further more each set of 4 generators is generated by a set of 4 tensor product of Pauli matrices by linear combination, the reason is due to fixing one of the Pauli matrices while permuting the other 2 Pauli matrices will produce 4 different tensor products with the same form because the identity matrices, $I$ and $\sigma_z$ are having real diagonal part. As for the factor of differentiating real  and imaginary generator come down to the contribution of off-diagonal part, $\sigma_x$ and $\sigma_y$. This further implies the similarity of change of basis. In addition to this, there was no contribution of $I\otimes I\otimes I$ component in any linear combination due to traceless properties of the generator of $SU(8)$. The results are especially important, as it shown that we could represent the Lie algebra $\mathfrak{su}(8)$ one-to-one in Pauli coordinates resembling change of coordinates which greatly reduced the amount of terms used for representation \citep{Nielsen_2006}, \citep{Dowling_2008}, \citep{Brandt_2009}. \\

In summary, we have computed the generator for $SU(8)$ Lie group is $\mathfrak{su}(8)$ Lie algebra, which is the generalized Gell-Mann matrices. We also have computed the tensor product of three Pauli matrices, and categorized them. We then compared both of them and computed the representation from one to another. It is shown to be a change of basis between generalized Gell-Mann matrices and tensor product of three Pauli matrices. Thus $\mathfrak{su}(8)$ can be represented in Pauli coordinates. 

\section*{Acknowledgments}\label{conclusion}

The authors wish to acknowledge the financial support provided through the IPS Grant, Project No. 9645700 Universiti Putra Malaysia.

\bibliographystyle{apa}
\bibliography{PauliSU}   

\end{document}